\documentclass[12pt,preprint]{aastex}
\usepackage{natbib}
\slugcomment{Accepted for publication in The Astrophysical Journal Letters}
\shorttitle{POLARIZATION OF L-DWARFS}
\shortauthors{SENGUPTA AND MARLEY }
\begin{document}

\title{OBSERVED POLARIZATION OF BROWN DWARFS SUGGESTS LOW SURFACE GRAVITY} 

\author{Sujan Sengupta}
\affil{Indian Institute of Astrophysics, Koramangala 2nd Block,
Bangalore 560 034, India; sujan@iiap.res.in}

\and

\author{Mark S. Marley}
\affil{NASA Ames Research Center, MS-245-3, Moffett Field, CA 94035,  
U.S.A.;
Mark.S.Marley@NASA.gov}

\begin{abstract}

Light scattering by atmospheric dust particles is responsible for the
polarization observed in some L dwarfs. Whether this polarization arises 
from an inhomogeneous distribution of dust across the disk or an 
oblate shape induced by rotation remains unclear. Here we argue that
the latter case is plausible and, for many L dwarfs, the more likely one.
Furthermore evolutionary models of mature field L dwarfs
predict surface gravities ranging from about 200 to 2500\, m\,s$^{-2}$
(corresponding to masses of $\sim$ 15 to 70 $M_{\rm Jupiter}$). Yet comparison
of observed spectra to available synthetic spectra often does not 
permit more precise determination of the surface gravity of individual
field L dwarfs, leading to important uncertainties in their properties.
Since rotationally-induced non-sphericity, which gives rise to non-zero
disk-integrated polarization, is more pronounced at lower gravities,
polarization is a promising low gravity indicator. Here we combine a
rigorous multiple scattering analysis with a self-consistent cloudy
atmospheric model and observationally inferred rotational velocities and find
that the observed optical  polarization can be explained if
the surface gravity of the polarized objects is about 300 m\,s$^{-2}$ or 
less, potentially providing a new method for constraining L dwarf masses.

\end{abstract}


\section{INTRODUCTION}

  Together with the lowest mass stars, brown dwarfs belong to the class 
of ultracool dwarfs whose relatively low atmospheric temperature and
high pressure results in clouds of refractory compounds that in turn
influence the entire atmosphere. The condensates (most abundantly iron
and Mg-silicate grains) form  near the atmospheric
temperature expected from chemical equilibrium as they efficiently deplete
the condensible species from the gas phase above the cloud 
\citep{ref4, ref5, ref6, ref7, ref8}.  As the objects cool over time,
the dust eventually settles down gravitationally below the visible atmosphere.

  The observed spectra and photometry of ultracool dwarfs have been compared
against predictions from models that incorporate the current understanding 
of atmospheric physics, chemistry, dynamics and most importantly cloud
processes \citep{ref9, ref10}. When combined with brown dwarf evolution tracks
\citep{ref11, ref12} these models can place fairly tight constraints on the
effective temperature, $T_{\rm eff}$, of dwarfs with quality spectra. However,
model fitting to date can leave the surface gravity of these objects
poorly constrained, often by up to an order of magnitude. Since the radii
of evolved brown dwarfs are only weakly dependent upon mass \citep{ref1, ref2},
surface gravity is nearly directly proportional to mass for mature field
L dwarfs older than several hundred million years. For such objects, 
surface gravities lying in the range of 200 to 2500 m\,s$^{-2}$ are expected
\citep{ref11}.  As very few brown dwarfs have known dynamical masses,
there has yet been few independent tests of the masses and temperatures
derived from the spectral fitting.
  
 Given the state of both observational and theoretical constraints on
dwarf surface gravity, an independent constraint on the surface gravity
of ultracool dwarfs is sorely needed.
There are some spectral indicators of low mass, particularly for early type
L dwarfs. \cite{Cruz09} identify weak alkaline absorption lines, and
differing strengths of metal oxides and hydrides as compared to 
typical L dwarf spectra as signs of low surface gravity by analogy to 
spectra of giant stars. The physical processes underlying these unusual
spectral features are as yet poorly understood and the value of surface
gravity at which they become apparent is uncertain. 
Imaging polarimetry can provide another independent metric of constraint for
surface gravity.

Linear polarization, almost certainly arising from dust scattering
\citep{ref13,sk05}, has been detected in the optical bands from a
good number of L dwarfs covering almost the entire range of spectral 
types L0--L8 \citep{ref14, ref15, ref16}.  Observations by \cite{ref14}
show that 25\% of L0--L3 dwarfs and 50\% of L3.5--L8 dwarfs in their sample
are intrinsically polarized while \cite{ref15} have detected
polarization from $15\pm9$\% of L0--L3 dwarfs and $43\pm17$\% 
of L3.5--L8 dwarfs in their samples.  In principle the observed polarization
could arise from the presence of magnetic field. However, radio, X-ray,
ultra-violet and H$\alpha$ observations point to a lack of magnetic activity
in mature field L dwarfs \citep{ref17}; magnetic field strengths in the range
100--1000G have been deduced from observations of 8.3 GHz radio emission from 
a few brown dwarfs. This implies that synchrotron processes will not lead to
significant linear polarization in the optical \citep{ref14}. Comparing the
small net linear polarization (of order a few times 0.01\%) detected from a 
sample of Ap stars \citep{ref18} with about 1 kG dipolar field at the surface, 
\cite{ref14} pointed out that the observed optical polarization of
ultracool dwarfs could not be explained by Zeeman splitting of atoms or
molecules. Furthermore  the warmer M dwarfs are found to be unpolarized
\citep{ref19}. M dwarfs have little or no atmospheric
dust although they should have stronger magnetic field \citep{ref20}.
So, we conclude dust scattering polarization is the most plausible physical 
process that can account for the observed polarization of L dwarfs.

Atmospheric dust can produce a net polarization if either  the dust is 
spatially inhomogeneous on large scales or the dust is homogeneous but
the disk  is oblate because of rotational distortion.  We argue here
that while the former mechanism cannot be ruled out, the latter is 
consistent with the current polarization  observations if the polarized objects
have relatively low gravity.  A rigorous polarization survey of L dwarfs
would distinguish between these mechanisms, providing either a new
method to probe dust cloud morphology or `weigh' the gravity of ultracool
dwarfs.

  \section{THE ATMOSPHERIC MODELS }

 In order to test if plausible, spatially uniform dust clouds can reproduce
the observed polarization of the L dwarfs, we employ a grid of 
one-dimensional atmosphere models \citep{ref4, ref21, ref22, ref11}
for specified $T_{\rm eff}$ and surface gravities $g$.
The atmosphere model parametrizes the efficiency of sedimentation of cloud
particles through a scaling factor $f_{\rm sed}$. 
For a fixed $T_{\rm eff}$, $g$ and $f_{\rm sed}$ the model uniquely predicts
the variation in mean particle size and particle number density through the
atmosphere which plays the crucial role in determining the scattering
polarization.  The atmosphere model employed here successfully reproduces
the spectra and photometry of a large number of L dwarfs at a wide range of
wavelengths covering near optical to mid-infrared regions as probed by
ground and space-based telescopes. Model fitting typically constrains the
effective temperature of an object of given spectral type within 100K and 
generally rule out the case for $f_{\rm sed} =1$ \citep{ref9, ref10}. 

The gas and dust opacity, the temperature-pressure profile and the dust
scattering asymmetry function averaged over each atmospheric pressure level
derived by the atmospheric code are used in a multiple scattering 
polarization code that solves the radiative transfer equations in vector
form to calculate the two Stokes parameter I and Q in a locally plane-parallel
medium \citep{ref24}. A combined Henyey-Greenstein-Rayleigh phase matrix
\citep{ref25} is used to calculate the angular distribution of the photons 
before and after scattering. Finally, the angle dependent I and Q are
integrated over the rotation-induced oblate disk of the object
by using a spherical harmonic expansion method and the degree of polarization is
taken as the ratio of the disk integrated polarized flux ($F_Q$) to the disk 
integrated total flux ($F_I$). The detail formalisms as well as the numerical
methods are provided in \cite{ref24}.

\section{ROTATION-INDUCED OBLATENESS }

 We employ the Darwin-Radau relationship \citep{ref26} for estimating the
rotation induced oblateness: 
\begin{eqnarray}\label{obl1}
f=1-\frac{R_p}{R_e}=
\frac{\Omega^2R}{g}\left[\frac{5}{2}\left(1-\frac{3K}{2}\right)^2+\frac{2}{5}
\right]^{-1}.
\end{eqnarray}
Here $R_e$ and $R_p$ are the equatorial and polar radii respectively, $\Omega$
is the spin angular velocity of the object and $K=I/(MR^2)$=0.261 for
polytropic index $n=1$ and 0.205 for $n=1.5$, $I$ being the moment of inertia.
Comparisons with detailed structure models
(D.~Saumon, private communication) show that irrespective of their age,
brown dwarf interiors can be adequately approximated by polytropes with
$1 < n < 1.3$ with the larger $n$ being appropriate for higher gravities.
As $n$ increases, the oblateness decreases for a given rotational velocity
and hence the degree of polarization decreases.
In the present work we consider $n=1$ and $n=1.5$ as two extreme cases.

\section{RESULTS AND DISCUSSION}

\subsection{Polarization from Oblate Dwarfs}

 Using the foregoing modeling approach we computed disk integrated 
polarization for a variety of model assumptions. The effect of varying 
surface gravity, viewing or inclination or projection angle $i$, $f_{\rm sed}$,
and rotation-induced oblateness on {\it I}-band (the bandpass at which most
of the data is available) polarization is presented in Fig.1 for a fixed 
effective temperature  $T_{\rm eff}=1800$K. The degree of polarization $p$
increases slowly with increasing oblateness and then increases rapidly for
oblateness greater than about 0.18. This is because for relatively
smaller oblateness the second harmonic in the spherical harmonic expansion 
is dominant, but as the oblateness increases the fourth and the sixth
harmonics contribute significantly increasing polarized flux $F_Q$. At 
the same time, higher harmonics become dominant and hence reduce the total 
flux $F_I$ (more reddening due to limb darkening). As a result, $p=F_Q/F_I$
increases rapidly. However, as the inclination angle decreases, the 
variation of polarization with respect to the oblateness changes noticeably.
All else being equal thicker clouds (smaller $f_{\rm sed}$) produce greater
polarization, especially when the surface gravity is high. 

    Fig.1 shows that the observed amount of L dwarf linear polarization
can be produced by dust scattering only if the oblateness is greater than 
about 0.18 irrespective of any allowed value of the parameters\footnote{The
stability limit oblateness for uniformly rotating polytropes with $n =1.0$
and 1.5 is  0.44 and 0.38 respectively (James 1964).}. In the
absence of a dust cloud, polarization at {\it I}-band is negligible for any
oblateness because Rayleigh scattering yields significant polarization only
at shorter wavelengths ({\it B}-band) \citep{ref24}.  As shown in Fig. 2, the
disk integrated degree of polarization $p$ remains almost the same
within the range of $T_{\rm eff}$ 1800--1300K roughly corresponding to 
spectral types L3--L8 and falls rapidly at higher $T_{\rm eff}$, where 
clouds form at lower pressure and are thinner. At $T_{\rm eff}<1800\,\rm K$,
clouds are found deeper in the atmosphere, are optically thicker, and produce
significant polarization. The transition from L to T dwarfs, i.e., from cloudy
to cloudless atmosphere occurs at about 1300K as the clouds dissipate
or settle below the photosphere.  Above $T_{\rm eff}\sim 2400$ K there are 
few condensates. Thus polarization is a marker for the presence of 
substantial cloud layers.

\subsection{Comparison to Observations}

 \cite{ref14} detected confirmed polarization from three L dwarfs,
marginal polarization from two and no polarization from three L dwarfs.
We consider the five confirmed and marginally polarized L dwarfs from 
this observation. \cite{ref15}  found confirmed polarization in
{\it I}-band from 9 L dwarfs out of 33 targets. Out of these 9 L dwarfs,
one (2MASS J1507--16) was also observed by \cite{ref16} who detected confirmed
polarization from three L dwarfs. \cite{ref16} detected polarization as 
high as $5.2\pm0.9$ \% in the {\it I}-band and $0.67\pm0.17$\% in the
{\it R}-band \citep{ref16} of 2MASS J1731+27. This object also shows a 
very high $H_{\alpha}$ equivalent width (-5.98) compared to the average 
value that is very small or zero. {\it Spitzer} IRAC observations exclude
a warm, but not a cold circumstellar disk \citep{ref16}. So, we exclude 
this object from consideration.

For each remaining 15 objects, we computed the rotational velocity required 
to produce sufficient oblateness to reproduce the observed polarization for
the $T_{\rm eff}$ of the object (based on its spectral type).  
The spectral types, adopted $T_{\rm eff}$, photometric variability, detected 
{\it I}-band polarization along with the associated errors and the projected
rotational velocity inferred from high resolution spectra as well as the 
same required to match the observed polarization are provided in Table 1.
$T_{\rm eff}$ for almost all objects is derived from the optical spectral 
type by using equation (4) of \cite{ref10}. For DENIS-P J2252--17, the infrared
spectral type is used as the optical spectral type is not known. $T_{\rm eff}$
calculated from optical and infrared spectral type differs by less than
100K for all objects except for 2MASS  J0141+18 which is a L1 object in 
optical but L4.5 object in infrared and so its effective temperature 
ranges between 2100K and 1550K as derived by using equations (4) and (3)
respectively of \cite{ref10}.

 The projected rotational velocity $V\sin(i)$ of a few L dwarfs showing
confirmed polarization is inferred observationally \citep{ref27, ref28, ref29}.
However, the projection angle $i$ is not known allowing a wide range of the 
values of $V$ that along with $g$ determines the oblateness. While a smaller
value of $i$ yields less polarization, it gives rise to higher rotational 
velocity for a fixed  $V\sin(i)$ and hence
more asymmetry. Taking the values of $V\sin(i)$ comparable to the observed 
values, we find that the models produce polarization comparable with that
observed in {\it I}-band only when the surface gravity less than 1000 ms$^{-2}$
and $i< 45^{\circ}$, leading to substantial disk asymmetry. Fig.2A shows
that all the five observational data points of \cite{ref14} can be fit 
well by setting  $i=30^{\circ}$ and $g=300$ ms$^{-2}$  with $V\sin(i)$ in range of
40 to 50 kms$^{-1}$ which is within or slightly higher than
the observed values. The observed {\it I}-band polarization of 2MASS J1507--16
fits well by using the observed $V\sin(i) = 27.2\,\rm km\,s^{-1}$ with
$g=300$ ms$^{-2}$ and $i=30^{\circ}$. 

Except for Kelu-1 and 2MASSW  J1412+16, the rotational velocity of most of
the polarized L dwarfs observed in \cite{ref15} is unknown. 
Fig. 2B shows that six out of the eight data from the \cite{ref15} sample
and one from \cite{ref16} -- 2MASS J1807+50 -- whose observed projected
rotational velocity is $76\,\rm km\,s^{-1}$ \citep{ref16, ref28}, 
can be fit if $V$ is as high as 90--$105\,\rm km\,s^{-1}$ and $i=90^{\circ}$ 
at which the polarization is maximum. As this corresponds to a rotation
period near an hour, a more likely explanation assuming homogeneous clouds
would be an even lower surface gravity for these objects. For example, 
if $V=80\,\rm km\,s^{-1}$, an equal amount of oblateness can be
achieved by lowering the surface gravity to 240--$175\,\rm m\,s^{-2}$.
Alternatively, other physical processes such as surface banding or 
inhomogeneity may give rise to such high polarization.  We note however
that the mean polarization value of \cite{ref15} could be high
because of the larger error bars owing to smaller telescope aperture than
in the other studies (F. Menard, private communication). 

Among the polarized objects, Kelu-1 is most likely a triple system 
with uncertain polarization contribution from the components. 2MASS J2244+20 
which shows polarization of $2.48\pm0.47$\% is extremeley red both in the
optical and near-infrared \citep{ref15}. Hence the high polarization
could be attributed to the presence of abnormally high amount of dust.
The high polarization of 2MASS  J1412+16, an L0.5 dwarf, while having small 
rotational velocity remains unexplained. Figure 2 shows that 
polarization does not alter drastically when the polytropic index is
increased from $n=1$ to $n=1.5$. For $n=1.5$, a slight increase in the
rotational velocity is needed in order to fit the observed data. However, 
for the range $1.0\leq n \leq 1.3$, the observed data can be fit without
altering the values of $V$, $g$ or $i$.

The range of $T_{\rm eff}$ and the value of $f_{\rm sed}$ adopted here
overlaps with that derived from spectral fit by \cite{ref10} for the three
common L dwarfs. All the data points in Fig. 2 can be fit with both
$f_{\rm sed}$=2 and 3 because for $g\leq 300$ ms$^{-2}$, the polarization
profile is not too sensitive to $3 \geq f_{\rm sed} \geq 2$ as implied by Fig. 1.

 Monte Carlo simulation of the field substellar mass function indicates
that objects in the 12-75 $M_{\rm Jupiter}$ mass range should greatly
outnumber lighter objects in the solar neighborhood and objects below
12-13 $M_{\rm Jupiter}$ are expected to constitute a modest fraction of 
field L dwarfs \citep{Bur04}.  Given the small sample sizes and lack of
uniform selection criteria  in the polarization surveys, it is premature to
draw any conclusion  from the relatively high fraction of low gravity
objects which we find.  Meanwhile the value of gravity below which the spectral
indicators identified by \cite{Cruz09} become apparent is not yet known,
thus the lack of `low gravity' spectral indicators in these objects is not
necessarily indicative.

\subsection{Polarization from Surface Inhomogeneities}

Inhomogeneous distribution of atmospheric dust and Jupiter like bands
\citep{ref19} may also produce detectable polarization.  
Whether or not the clouds of L dwarfs are homogenous or patchy remains
an open issue.  Existing spectral models assume spatially uniform dust
clouds and generally accurately reproduce observed L dwarf spectra
\citep{ref10}. Surface inhomogeneities can produce photometric variability
and 40-50\% of L dwarfs are found to be variable without any periodicity 
\citep{Gel02, Koen03}. However, 2MASS J2224-01, 2MASS J1108+68, 2MASS J1658+70
\citep{Gel02} and 2MASS J1048+01 \citep{Koen03} are all variable L dwarfs with
no detectable polarization implying that inhomogeneities do not always 
produce significant polarization. On the other hand, 2MASS  J1412+16,
2MASS J0036+18 \citep{Gel02}, and 2MASS J1507-16 \citep{Koen03}
are not variable implying a lack of large scale surface inhomogeneity (or a 
particularly favorable morphology and viewing geometry) but all of them  are
polarized. Another example is  2MASSW  J1048+01 which is a variable object
with comparatively low projected velocity ($V\sin i=17\,\rm km\,s^{-1}$)
\citep{ref28}. If inhomogeneity produces detectable amounts of 
polarization then this object should show polarization irrespective of its
low rotational velocity but it is unpolarized.  

An alternative hypothesis is that some dwarfs exhibit a uniform, banded
appearance. Such objects might still be polarized but not be variable. 
We conclude that unlike the case for oblateness induced polarization--which
naturally explains both the magnitude and variation with spectral type of
polarization (Figure 2)--surface inhomogeneities require
reliance on special cloud morphologies and viewing angles.  Nevertheless
both mechanisms likely play a role in producing polarization in some objects.

\section{CONCLUSIONS}

We have found that model L dwarf atmospheric structures which generally well
reproduce the spectra of known objects, predict full disk polarization of L
dwarfs comparable to the values observed on some objects if the dwarfs are
substantially oblate.  Because the degree of oblateness varies inversely
with gravity (Eq. 1), this mechanism requires fairly low surface gravities
($g\sim 300\,\rm m\, s^{-2}$).  Thus if oblateness is the primary 
mechanism by which L dwarfs become polarized, then polarization
is a marker for low gravity.  While some spectral indicators of low gravity
have been identified in the literature \citep[e.g.][]{Cruz09},
the actual gravity at which they become prominent has yet to be established. 
Thus polarization may serve as an indicator of moderately low L-dwarf gravity,
at least for rapidly rotating dwarfs.

A well constructed survey of a sample of field L dwarfs for polarization,
$V\sin(i)$, variability, and spectral gravity indicators
would test more definitively which mechanism (oblateness or surface
inhomogeneities) is primarily responsible for L dwarf polarization.
Once established, polarization would serve as a new constraint on the
properties of newly discovered objects. However, this conclusion relies
on the assumption that the adopted dust model describes the cloud
distribution correctly.    

 Finally, we predict (Fig. 3) that the degree of polarization at {\it J}-band
is comparable with that of {\it I}-band but is reduced at {\it H}- and
{\it K}- bands.  Again, in the infrared we expect a detectable 
amount of polarization only if the surface gravity is about 
$300\,\rm m\,s^{-2}$ or less.

\section{Acknowledgements}

We thank M. Cushing, F. Menard, D. Saumon and I. Baraffe for useful 
discussions and suggestions. We also thank the referee for 
critical comments and useful suggestions.

\clearpage
\begin{figure}
\includegraphics[angle=0.0,scale=.70]{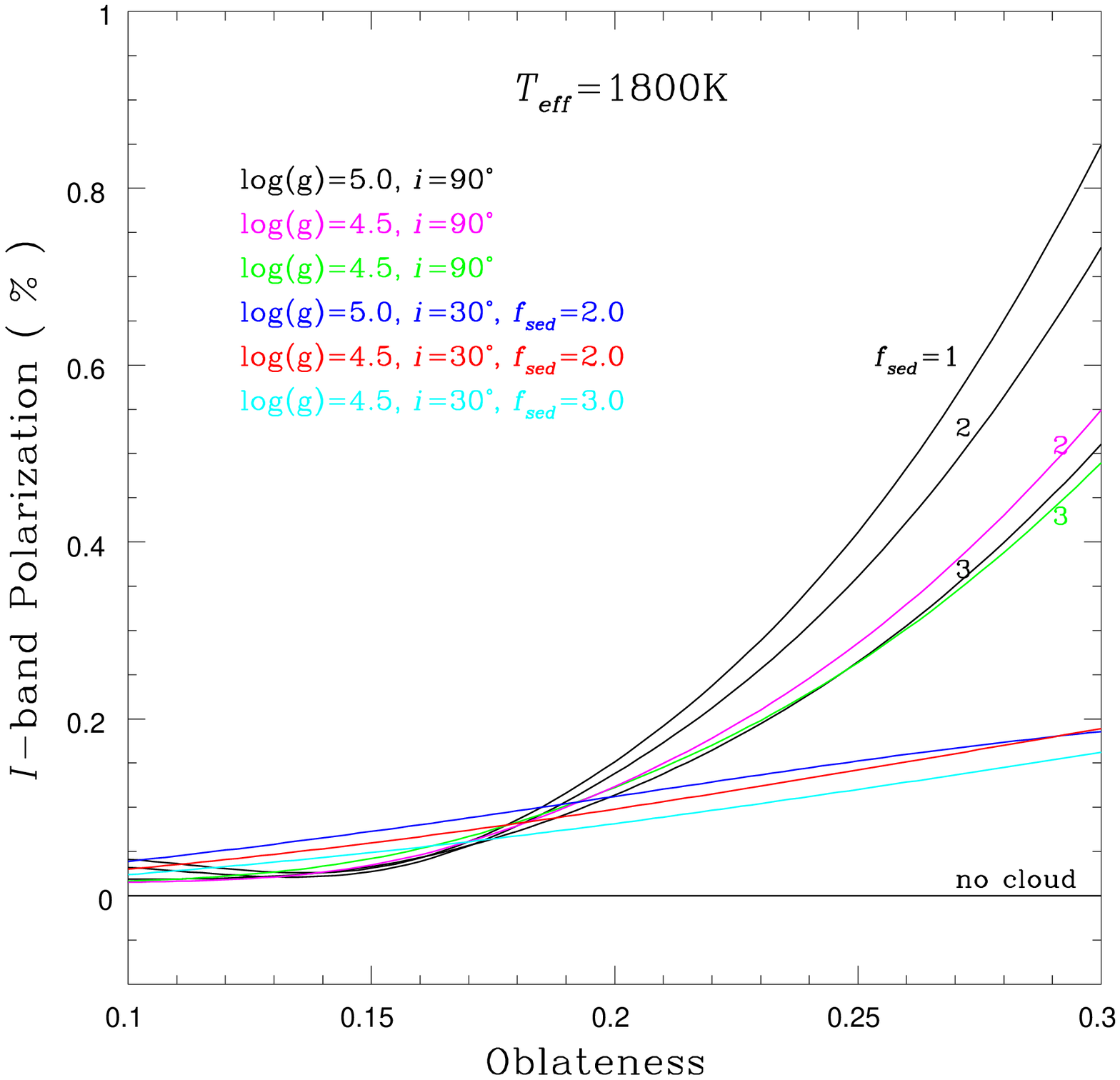}
\caption{ Percentage degree of linear polarization calculated for the 
{\it I}-band as a function of oblateness. The numbers near the curves
correspond to the value of the sedimentation efficiency parameter
$f_{\rm sed}$.  The polytropic index $n=1$.
\label{fig1}}
\end{figure}

\begin{figure}
\includegraphics[angle=0.0,scale=.70]{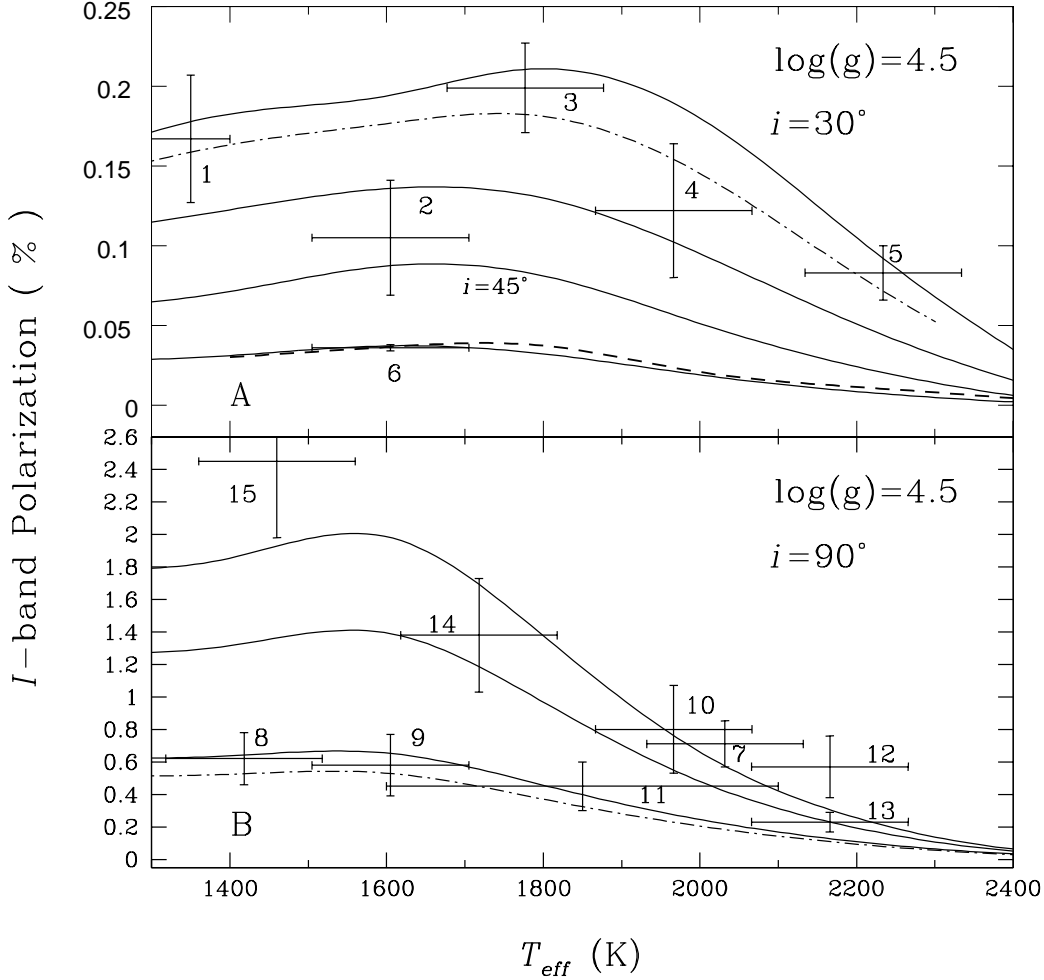}
\caption{ Model fits of the observed {\it I}-band polarization.
The vertical error bars are observational errors and the horizontal ones
are the spread of effective temperature for a particular spectral type.
The numbers near the error bars correspond to the objects as listed in
table~1. For all the cases $f_{\rm sed}=2$. (A) The solid lines represent
model with surface gravity $g=300$ ms$^{-2}$ and $n=1$. From top to bottom
they represent model with $V\sin(i)=48, 41, 48,$ and $27.2\,\rm
km\,s^{-1}$ respectively.  The dot-dash line represents model with $n=1.5$ and
$V\sin(i)=50\,\rm km\,s^{-1}$. The dash
line represents that with $g=1000\,\rm m\,s^{-2}$, $n=1$ and $V\sin(i)=48\,\rm
km\,s^{-1}$. For all the cases except the one marked otherwise, $i=30^{\circ}$.
(B) same as (A) but $i=90^{\circ}$ and $g=300$ ms$^{-2}$ for all the
cases. From top to bottom, the solid lines represent model with
$V\sin(i)=105, 100$ and $90\,\rm km\,s^{-1}$ respectively.
The dash-dot line represents model with $n=1.5$ and $V\sin(i)=96$ kms$^{-1}$.
\label{fig2}}
\end{figure}

\begin{figure}
\includegraphics[angle=0.0,scale=.70]{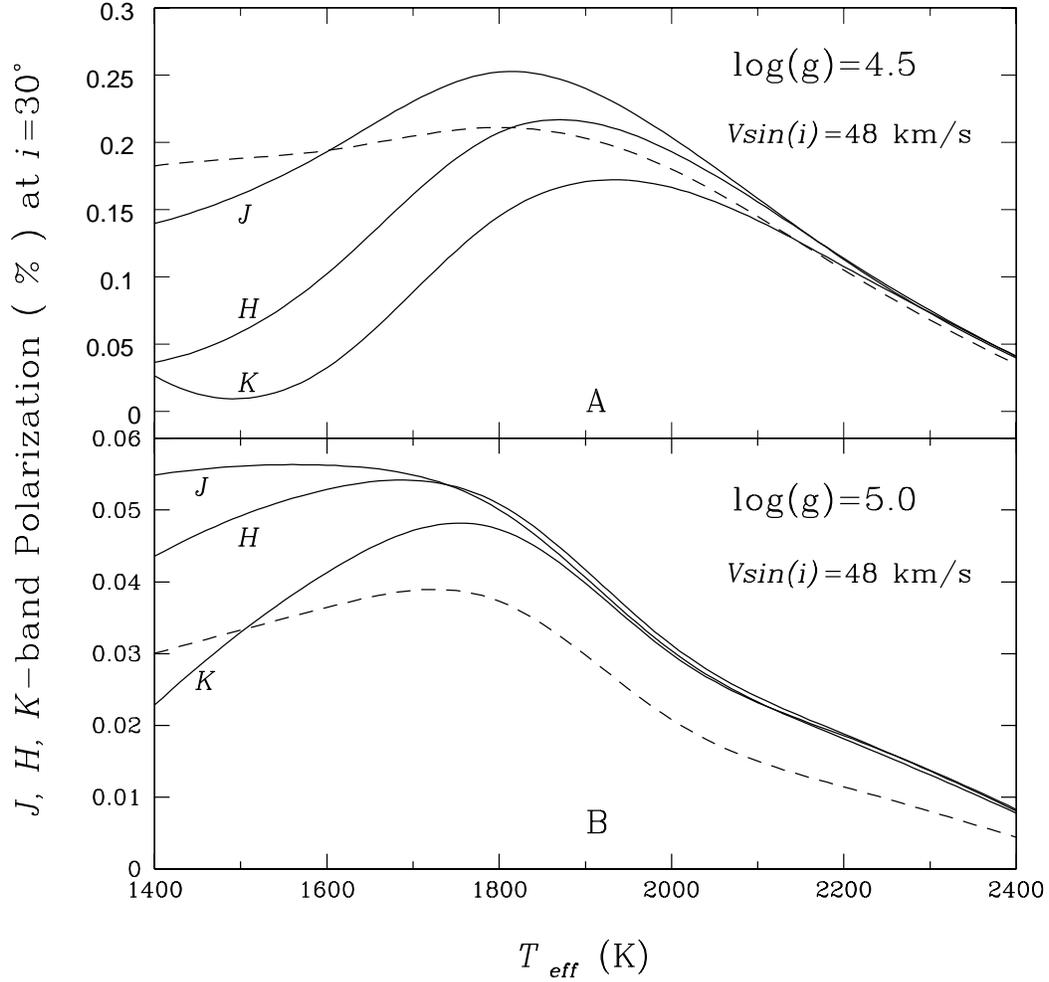}
\caption{Predicted {\it J}-, {\it H}-, and {\it K}-band polarization of
L dwarfs at different $T_{\rm eff}$ and for different surface gravities. 
Only the model that fits the observed {\it I}-band polarization of a few 
L dwarfs is presented. Broken lines represent the polarization for {\it I}-band.
\label{fig3}}
\end{figure}

\clearpage
\begin{table}
\begin{center}
\caption{Observed and derived quantities of polarized L dwarfs.}
\begin{tabular}{cccccccc}
\tableline\tableline
Serial &  Discovery &  Spectral & Estimated &
Photometric   & $p$\% & $\sigma$ \% & $V\sin(i)$ \\
No. & name & type  & $T_{\rm eff}$(K) & variability & & & (kms$^{-1}$) \\
& & (Optical) & &  & \\
\tableline 
1 & DENIS-P & L8 & 1340$\pm$ 50 & Yes  & 0.167 & $\pm$
0.04 & 40.8$\pm$ 8 \tablenotemark{a} \\
& J0255-4700 & & (1200-1300) &  & & & (41-48) \\
2 & LHS 102B &  L5 & 1605 $\pm$ 100 & -- & 0.105 & $\pm$ 0.036 & 32.5\tablenotemark{a} (41) \\
3 & 2MASS  & L3.5 & 1780$\pm$ 100 & No & 0.199 & $\pm$ 0.028 & 
45\tablenotemark{b} \\
& J0036+1821 & & (1700-1800) & & & & (48) \\
4 & DENIS-P  & L2 & 1966.5 $\pm$ 100 & -- & 0.122 & $\pm$ 0.042 & --
 \\
& J2036-1306 & & & & & & (41-48) \\
5 & DENIS-P & L0 & 2234 $\pm$ 100 & -- & 0.083 & $\pm$ 0.017 & --\\
& 2000-7523 & & & & & & (48) \\
6 & 2MASS  & L5 & 1605.5 $\pm$ 100  & No & 0.036 & 
0.0  & 27.2 \tablenotemark{c} \\
& J1507-1627 & & (1600-1700) & & & & (27.2) \\
7 & 2MASS & L1.5 & 2100 $\pm$ 100 & -- & 0.711 & $\pm$ 0.142 & 76\tablenotemark{b} \\
& J1807+5015 & & & & & & (105) \\
8 & Denis-P & L7.5(IR) & 1419 $\pm$ 100 & -- & 0.62 & $\pm$ 0.16 &
-- \\
& J2252-1730 & & & & & & (90) \\
9 & 2MASS  & L5 & 1605 $\pm$ 100 & -- & 0.58 & $\pm$ 0.19 & -- \\
& J0144-0716 & & & & & & (90) \\ 
10 & Kelu-1 & L2 & 1966.5 $\pm$ 100 & Yes & 0.8 & $\pm$ 0.27 &
60\tablenotemark{a} \\
& & & & & & & (105) \\
11 & 2MASS  & L1/L4.5 & 1850 $\pm$ 250 & -- & 0.45 & $\pm$ 0.15 & --
\\
& J0141+1804 & & & & & & (90-105) \\
12 & 2MASS  & L0.5 & 2166 $\pm$ 100 & No & 0.57 & $\pm$ 0.19 & 19 \tablenotemark{b} \\
& J1412+1632 & & & & & & (--) \\
13 & 2MASS & L0.5 & 2166 $\pm$ 100 & -- & 0.23 & $\pm$ 0.06 & --\\
& J1707+4301 & & & & & & (105) \\
14 & 2MASS & L4.0 & 1718 $\pm$ 100 & -- & 1.38 & $\pm$ 0.35 & -- \\
& J2158-1550 & & & & & & (100) \\
15 & 2MASS & L6.5 & 1460 $\pm$ 100 & very red & 2.45 & $\pm$ 0.47 & --  \\
& J2244+2043 & & & & & & ($\sim105$) \\
\tableline
\end{tabular}
\tablenotetext{a}{\cite{ref27}}
\tablenotetext{b}{\cite{ref28}}
\tablenotetext{c}{\cite{ref29}}
\tablecomments
{In column 4, the numbers inside brackets are $T_{\rm eff}$  
derived from synthetic spectra by \cite{ref10} . $p$ is the observed
{\it I}-band mean polarization and $\sigma$ is the associated error. 
For objects 1-5, $p$ and $\sigma$ are taken from \cite{ref14}, 6-7
from \cite{ref16}, 8-15 from \cite{ref15}.  In the last column, the
numbers inside brackets are the values of the projected rotational velocity
required to achieve the observed polarization while that outside brackets are 
the same inferred from high resolution spectroscopy.}
\end{center}
\end{table}

\end{document}